# A Comparison Between Human and Generative AI Decision-Making Attributes in Complex Health Services

Short title: Comparing Human and Generative AI Decision-Making Attributes

Nandini Doreswamy[1, 2], MBBS, MS, MBA (ORCID 0000-0002-8467-762X)

[1] Southern Cross University, Lismore, New South Wales, Australia [2]National Coalition of Independent Scholars, ndoreswamy@outlook.com

Louise Horstmanshof, PhD, MOrgPsych, MAPs (ORCID 0000-0002-0749-1231)

Southern Cross University, Lismore, New South Wales, Australia, louise.horstmanshof@scu.edu.au

**Abstract**

A comparison between human and Generative AI decision-making attributes in complex health services is a knowledge gap in the literature, at present. Humans may possess unique attributes beneficial to decision-making in complex health services such as health policy and health regulation, but are also susceptible to decision-making flaws. The objective is to explore whether humans have unique, and/or helpful attributes that contribute to optimal decision-making in complex health services. This comparison may also shed light on whether humans are likely to compete, cooperate, or converge with Generative AI. The comparison is based on two published reviews: a scoping review of human attributes [1] and a rapid review of Generative AI attributes [2]. The analysis categorizes attributes by uniqueness and impact. The results are presented in tabular form, comparing the sets and subsets of human and Generative AI attributes. Humans and Generative AI decision-making attributes have complementary strengths. Cooperation between these two entities seems more likely than pure competition. To maintain meaningful decision-making roles, humans could develop their unique attributes, with decision-making systems integrating both human and Generative AI contributions. These entities may also converge, in future.

## 1 INTRODUCTION

From the middle of the twentieth century, health policy and health regulation developed as domains distinct from clinical healthcare, requiring new systems of knowledge and expertise [3]. Today, these complex health services are often global in scope, requiring nuanced decision-making with far-reaching consequences, as evidenced by the COVID-19 pandemic. Generative AI has increased the complexity of decision-making in these services. Therefore, it is timely to compare the decision-making attributes of humans and Generative AI in this context.

It is possible that humans may have certain unique, helpful attributes that influence health policy decision-making for the better, when compared to Generative AI. If so, these attributes may help humans continue in meaningful decision-making roles in an AI-driven future. For example, cognitive complexity, flexibility, processing, and performance, and the ability to apply these internally and externally, are attributes that may well enhance the human value proposition. Humans may offer a holistic and intuitive approach to decision-making [4] that could present a competitive advantage in the future.

On the other hand, human beings also have attributes that may detract from sound judgement and decision-making. For example, escalation of commitment is a human characteristic that results in a tendency to persist with decisions to allocate time, money, or even identity, to courses of action that are either failing, or have already failed [5-8]. This is closely related to the sunk cost fallacy, a negative affect [9] associated with the tendency to commit more time, effort, and money to an outcome, even when that outcome is increasingly negative [5, 8, 9]. Thus far, sunk cost fallacy has only been observed in human beings [10].

Humans still dominate decision-making in complex health services, but Generative AI is showing ever-increasing utility and influence. Complex health services are beginning to incorporate sophisticated Generative-AI-based decision support systems [11]. It is only a matter of time before Generative AI begins to drive or dominate decision-making in these services. Therefore, this research is timely and essential.

There is an urgent need to determine whether humans can continue in meaningful decision-making roles in a future dominated by Generative AI. There is a dearth of literature on comparisons between human and Generative AI attributes that influence the decision-making of each of these entities in the context of complex health services.

### 1.1 Significance of This Comparison

This research may inform leaders, managers, and decision makers about the likelihood of continuing in meaningful decision-making roles in a Generative-AI-driven future in complex health services. This comparison may shed light on whether humans have one or more unique and/or helpful attributes that could influence decision-making in this context, that may allow them to continue to play a meaningful role in a future dominated by Generative AI. This comparison may also address the question of whether humans will compete, cooperate, or converge with Generative AI to continue in decision-making roles in this context.

### 1.2 Rationale and Objectives

The objective of this comparison is to investigate whether humans have one or more unique, and/or helpful attributes that influence decision-making in complex health services, that may enable them to continue as decision makers in these heath domains. This comparison may also shed light on whether humans are likely to compete, cooperate, or converge with Generative AI.

## 2 METHODS

This is a narrative comparison between two sets of attributes—human and Generative AI—in complex health services, that were reported in the literature. The comparison is based on two reviews.

- A scoping review that identified and mapped attributes that influence human decision-making in complex health services, as reported in the literature [1].
- A rapid review that identified and mapped Generative AI attributes in the same context [2].

Figure 1 shows the attributes that influence human decision-making in complex health services, as reported in the literature.

![Figure 1 treemap]

Figure 1: Human attributes (n=45) that influence decision-making in complex health services: frequency of mentions in included papers [1].

Figure 2 shows the attributes that influence Generative AI decision-making in complex health services, as reported in the literature.

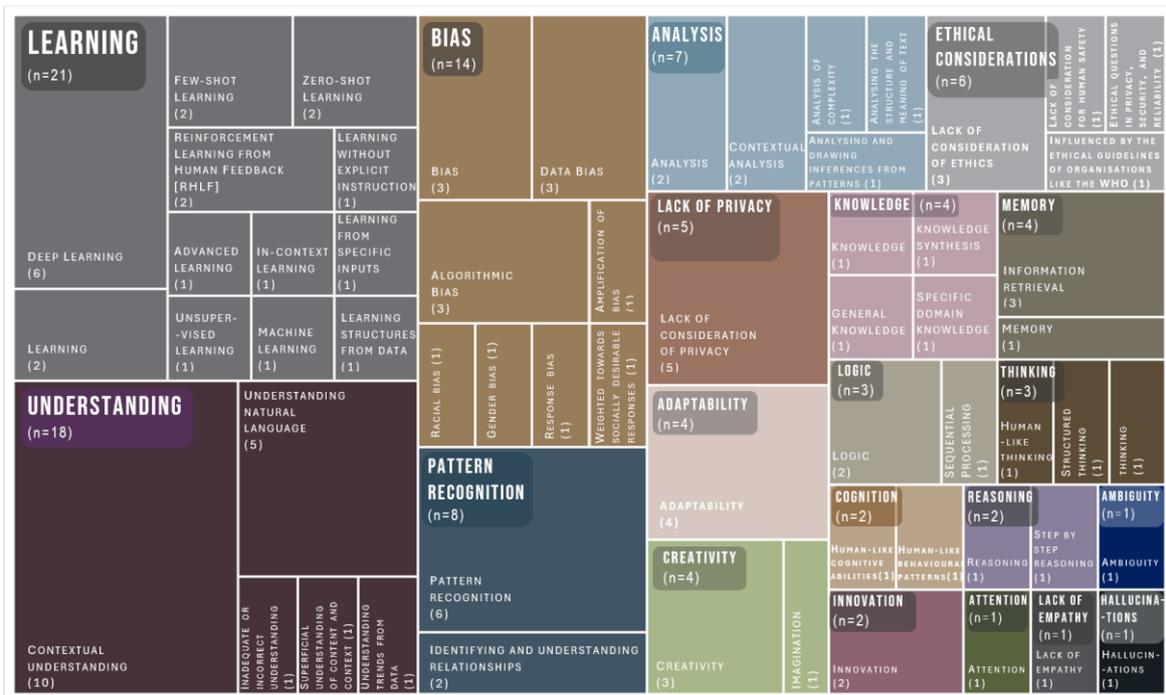

Figure 2: Attributes of Generative AI that influence its decision-making in complex health services and frequency of attributes (n=111) mentioned in the articles included [2].

Each group of attributes (human and Generative AI) was viewed as a set. These sets were each divided into two sub-sets:

1. Attributes that are unique and non-unique to each entity: humans and Generative AI.

2. Attributes based on their impact on effective decision-making: those that are helpful, those that are detrimental, and those dependent on context.

A narrative comparison between these sets and sub-sets was undertaken to determine whether humans may have one or more unique attributes that exert an influence on appropriate decision-making. The results are presented in Table 1.

Table 1: A Comparison Between Human and Generative AI attributes that influence decision-making in complex health services

| # | Human Attributes (n=45) | Helpful, Detrimental, or Context-dependent | Generative AI Attributes (20) | Helpful, Detrimental, or Context-dependent | Generative AI Attribute Sub-category | Are the Attributes a Match, a Mirror Image, Related, or Unique? |
|---|---|---|---|---|---|---|
| 1 | Rationality | Helpful (+) | Analysis | Helpful (+) | + Analysis<br>+ Contextual analysis<br>+ Analyzing and drawing inferences from patterns<br>+ Analyzing the structure and meaning of text<br>+ Analysis of complexity | Related |
|   |   |   | Logic | Helpful (+) | + Logic<br>+ Sequential processing | Related |
|   |   |   | Reasoning | Helpful (+) | + Reasoning<br>+ Step by step reasoning | Related |
| 2 | Expertise | Helpful (+) | Learning | Helpful (+) | + Deep learning<br>+ Few-shot learning<br>+ Learning<br>+ Reinforcement learning from human feedback (RLHF)<br>+ Zero-shot learning<br>+ Advanced learning<br>+ In-context learning<br>+ Learning from specific inputs<br>+ Learning structures from data<br>+ Learning without explicit instruction<br>+ Machine learning<br>+ Unsupervised learning | Related |
|   |   |   | Knowledge | Helpful (+) | + General knowledge<br>+ Knowledge<br>+ Knowledge synthesis<br>+ Specific domain knowledge | Related |
| 3 | Phronesis (ability to apply personal, specialist, or experiential knowledge) | Helpful (+) | Understanding | Context-dependent (±) | + Contextual understanding<br>+ Understanding natural language<br>+ Understanding trends from data<br>- Inadequate or incorrect understanding<br>- Superficial understanding of content and context | Related |
|   |   |   | Pattern recognition | Helpful (+) | + Pattern recognition<br>+ Identifying and understanding relationships | Related |
| 4 | Morality | Helpful (+) |   |   |   | Unique to humans |
| 5 | Cognitive bias (including social bias) | Detrimental (-) | Bias | Detrimental (-) | - Algorithmic bias<br>- Bias<br>- Data bias |   |

| # | Human Attributes (n=45) | Helpful, Detrimental, or Context-dependent | Generative AI Attributes (20) | Helpful, Detrimental, or Context-dependent | Generative AI Attribute Sub-category | Are the Attributes a Match, a Mirror Image, Related, or Unique? |
|---|---|---|---|---|---|---|
| | | | | | - Amplification of bias<br>- Gender bias<br>- Racial bias<br>- Response bias<br>- Weighted towards socially desirable responses | |
| | | | Cognition | Helpful (+) | + Human-like behavioral patterns<br>+ Human-like cognitive abilities | Related |
| 6 | Collective understanding | Helpful (+) | Understanding | Context-dependent (±) | + Contextual understanding<br>+ Understanding natural language<br>+ Understanding trends from data<br>- Inadequate or incorrect understanding<br>- Superficial understanding of content and context | Related |
| 7 | Compassion | Helpful (+) | | | | Unique to humans |
| 8 | Dialogical reasoning and dialectical thinking | Helpful (+) | Reasoning | Helpful (+) | + Reasoning<br>+ Step by step reasoning | Related |
| | | | Thinking | Helpful (+) | + Human-like thinking<br>+ Structured thinking<br>+ Thinking | Related |
| | | | Innovation | Helpful (+) | + Innovation | Related |
| | | | Creativity | Helpful (+) | | Related |
| 9 | Emotion | Context-dependent (±) | | | | Unique to humans |
| 10 | Ethics | Helpful (+) | Ethical considerations | Context-dependent (±) | + Influenced by the ethical guidelines of organisations like the World Health Organization (WHO)<br>- Lack of consideration for ethics<br>- Lack of consideration for human safety<br>- Ethical questions in privacy, security, and reliability | Related / Mirror image |
| | | | Lack of privacy | Detrimental (-) | - Lack of consideration of privacy | Related / Mirror image |
| 11 | Heuristics | Helpful (+) | | | | Unique to humans |
| 12 | Sense-making | Helpful (+) | Pattern recognition | Helpful (+) | + Pattern recognition<br>+ Identifying and understanding relationships | Related |
| 13 | Values | Helpful (+) | | | | Unique to humans |
| 14 | Thoughtfulness | Helpful (+) | | | | Unique to humans |
| 15 | Cognitive limitations of memory | Detrimental (-) | Memory | Helpful (+) | + Information retrieval<br>+ Memory | Mirror image |
| 16 | Narrative reasoning | Helpful (+) | Reasoning | Helpful (+) | + Reasoning<br>+ Step by step reasoning | Related |
| 17 | Logic | Helpful (+) | Logic | Helpful (+) | + Logic<br>+ Sequential processing | Match |

| # | Human Attributes (n=45) | Helpful, Detrimental, or Context-dependent | Generative AI Attributes (20) | Helpful, Detrimental, or Context-dependent | Generative AI Attribute Sub-category | Are the Attributes a Match, a Mirror Image, Related, or Unique? |
|---|---|---|---|---|---|---|
| 18 | Adaptation | Helpful (+) | Adaptability | Helpful (+) | + Adaptability | Match |
| 19 | Empathy | Helpful (+) | Lack of empathy | Detrimental (-) | - Lack of empathy | Mirror image |
| 20 | Intuition | Helpful (+) | | | | Unique to humans |
| 21 | Beliefs | Context-dependent (±) | | | | Unique to humans |
| 22 | Humility | Helpful (+) | | | | Unique to humans |
| 23 | Aspiration | Helpful (+) | | | | Unique to humans |
| 24 | Flexibility | Helpful (+) | Adaptability | Helpful (+) | + Adaptability | Related |
| 25 | Inflexibility | Context-dependent (±) | | | | Unique to humans |
| 26 | Risk perception | Context-dependent (±) | | | | Unique to humans |
| 27 | Inertia of actions | Context-dependent (±) | | | | Unique to humans |
| 28 | Holistic approach | Helpful (+) | | | | Unique to humans |
| 29 | Present-time based preferences | Context-dependent (±) | | | | Unique to humans |
| 30 | Give and take | Helpful (+) | | | | Unique to humans |
| 31 | Critical reflection | Helpful (+) | | | | Unique to humans |
| 32 | Discretion | Context-dependent (±) | | | | Unique to humans |
| 33 | Fear | Context-dependent (±) | | | | Unique to humans |
| 34 | Ability to deal with complexity | Helpful (+) | | | | Unique to humans |
| 35 | Maximization | Helpful (+) | | | | Unique to humans |
| 36 | Engagement | Helpful (+) | | | | Unique to humans |
| 37 | Relationships and alliances | Helpful (+) | | | | Unique to humans |
| 38 | Social considerations | Helpful (+) | | | | Unique to humans |
| 39 | Political considerations | Helpful (+) | | | | Unique to humans |
| 40 | Framing | Helpful (+) | | | | Unique to humans |
| 41 | Interaction | Helpful (+) | | | | Unique to humans |
| 42 | Monetary motivation | Context-dependent (±) | | | | Unique to humans |
| 43 | Deliberation | Helpful (+) | | | | Unique to humans |
| 44 | Satisfaction | Helpful (+) | | | | Unique to humans |
| 45 | Preferences | Context-dependent (±) | | | | Unique to humans |
| | | | Ambiguity | Detrimental (-) | - Ambiguity (1) | Unique to Generative AI |
| | | | Attention | Helpful (+) | + Attention (1) | Unique to Generative AI |
| | | | Hallucinations | Detrimental (-) | - Hallucinations (1) | Unique to Generative AI |

## 3 DISCUSSION

Of the 45 human attributes identified in the scoping review [1], 26 (57.8%) were deemed helpful for effective decision-making, while two (4.4%) were considered detrimental, and 17 (37.8%) were context-dependent. In contrast, from the 20 major Generative AI attributes identified, 14 (70%) were classified as helpful, three (15%) as detrimental, and three (15%) as context-dependent. This preliminary quantitative analysis suggests that both entities possess predominantly helpful attributes for decision-making in complex health services, though in varying proportions and with different qualitative characteristics.

## 3.1 Uniqueness of Human Attributes

*3.1.1 Nature and Significance of Unique Human Attributes*

Humans have a number of unique and beneficial attributes that could potentially secure their role as decision makers in an AI-driven future. In this comparison, 25 attributes (over 55% of all human attributes) were found to be unique to humans, with no corresponding attributes in Generative AI. Significantly, 17 of these unique attributes were classified as beneficial to effective decision-making. These include morality, compassion, heuristics, values, thoughtfulness, intuition, humility, aspiration, a holistic approach, the ability to deal with complexity, maximization, engagement, relationships and alliances, framing, interaction, deliberation, and satisfaction.

The uniqueness of human attributes largely centers around contextual understanding, the interpersonal domain, and morality. For instance, compassion—the ability to recognize suffering and the motivation to alleviate it—is a fundamentally human characteristic that influences healthcare decision-making in ways that transcend logic alone. Similarly, morality, values, and ethical frameworks provide humans with internally consistent guidance that helps navigate potential ethical dilemmas in complex health services.

As Jarrahi [4] suggested, humans offer a holistic and intuitive approach to decision-making, which could present a competitive advantage over Generative AI. Attributes such as intuition, a holistic approach, and the ability to deal with complexity allow humans to consider multiple dimensions simultaneously, incorporate tacit knowledge, and recognize patterns even with incomplete information. Despite its computational advantage, these capabilities remain a challenge for Generative AI.

*3.1.2 Context-dependent and Potentially Detrimental Human Attributes*

Not all unique human attributes are universally beneficial. This comparison has identified several context-dependent attributes, such as emotion, beliefs, inflexibility, risk perception, inertia of action, present-time based preferences, discretion, fear, and monetary motivation, that may enhance or impede effective decision-making, depending on the context. For example, emotions such as compassion and empathy are beneficial, but the emotional stress and burnout seen in healthcare professionals in the COVID-19 pandemic were detrimental to effective decision-making [3].

On the other hand, escalation of commitment and sunk cost fallacy as potentially detrimental to making good decisions. While these specific terms were not explicitly identified as separate attributes in our analysis, they relate to several identified attributes including inflexibility, inertia of action, and present-time based preferences. These tendencies, as noted by Arkes and Blumer [5] and Thaler [8] can lead humans to persist with failing courses of action. This vulnerability is not shared by Generative AI, which can more readily recalibrate when presented with new evidence, without emotional or identity-based attachments to previous decisions.

## 3.2 Unique Attributes of Generative AI

Based on data extracted from the reviews [1, 2] three attributes are unique to Generative AI: attention, which is helpful to effective decision-making, and ambiguity and hallucinations, which are detrimental. Generative AI hallucinations result in content that appears plausible, but is factually incorrect or unsupported by evidence. Hallucinations present a significant concern in contexts such as decision-making in health policy or health regulation, where accuracy is paramount. Ambiguity in Generative AI responses could lead to misinterpretation and potentially harmful decisions in complex health services. Attention allows Generative AI to focus on relevant sections of data, enabling it to process and prioritize information, and sift through vast amounts of literature, including legal precedents, patient records, and clinical and policy guidelines, to identify appropriate information. This is a significant advantage in the information-dense environments of complex health services.

## 3.3 Overlapping Attributes: Related, Matched, or Mirror Images

The majority of Generative AI attributes (17 out of 20, or 85%) overlap with human attributes in several ways. These relationships were categorized as follows:

1. **Matched**: Both humans and Generative AI have logic and adaptability, which are fundamental cognitive processes [12, 13].

2. **Related**: Many attributes, including rationality/analysis, expertise/learning, phronesis/understanding, cognitive bias/bias, dialogical reasoning/reasoning, narrative reasoning/reasoning, and flexibility/adaptability share conceptual foundations while manifesting differently in humans and Generative AI.
3. **Mirror images**: Some attributes are mirror images in humans and Generative AI. Humans are capable of empathy, whereas Generative AI lacks this attribute. Humans are capable of ethical considerations, whereas Generative AI, although influenced by the ethical guidelines of international bodies like the World Health Organization (WHO), usually lack consideration for ethics, privacy, and human safety. On the other hand, humans have cognitive limitations of memory, whereas Generative AI does not have such limitations. These mirror-image relationships suggest potential complementarity, where the strengths of one entity might compensate for the weaknesses of the other.

**3.4 Implications for Decision-Making in Complex Health Services**

*3.4.1 Potential for Human-AI Complementarity*

The pattern of attributes identified suggests that humans and Generative AI possess complementary strengths and weaknesses. For example, while humans demonstrate empathy, holistic thinking, and moral reasoning, they suffer from cognitive bias and limitations of memory. Conversely, Generative AI excels in information retrieval, pattern recognition, and memory but struggles with empathy and can have hallucinations and ambiguous outputs.

This complementarity suggests that cooperation, rather than competition between humans and Generative AI may offer an effective approach to decision-making in complex health services, in the future. As Prince, Barrett and Oborn [14] and Paletz, Bogue, Miron-Spektor and Spencer-Rodgers [15] noted, the integration of diverse cognitive approaches often yields superior outcomes, when compared with homogeneous thinking. Cooperation between humans and Generative AI can lead to cognitive diversity that is beneficial to making effective decisions. For instance, humans could provide ethical judgment, contextual understanding, and interpersonal skills, while Generative AI sifts through substantial amounts of data, recognizes patterns across data sets, and rapidly retrieves relevant information.

*3.4.2 Areas of Competitive Advantage for Humans*

Despite the computational advantages of Generative AI, humans maintain several domains of competitive advantage based on beneficial attributes unique to them.

1. **Ethical and moral reasoning**: Human capacity for morality, compassion, values, and ethical reasoning provides a foundation for navigating complex situations and paradigms. For example, Generative AI algorithms may fall short in a pandemic, where healthy policy decisions at the local, national, and global scale require many attributes that are uniquely human.
2. **Interpersonal capabilities**: Attributes like empathy, relationships and alliances, interaction, and social considerations enable humans to build trust, communicate effectively with key stakeholders and colleagues, and understand the social dimensions of decision-making in complex health services. Generative AI lacks these attributes. While it can communicate, it may struggle to build trust.
3. **Contextual understanding**: The human ability to understand cultural, political, and institutional contexts through attributes like framing, political considerations, and a holistic approach allows for nuanced decision-making that accounts for local circumstances and stakeholder dynamics. Generative AI may, perhaps, lack the depth of understanding required and miss the subtleties in these contexts.
4. **Metacognitive processes**: The human capacity for critical reflection, thoughtfulness, deliberation, and humility enables them to recognize the limitations of their knowledge and adjust their decision-making processes accordingly. Generative AI lacks these uniquely human attributes and, therefore, lacks the insight needed to adjust its decisions appropriately.

*3.4.3 Challenges for Human Decision Makers*

While humans possess many unique and beneficial attributes, they also face significant challenges in competing with Generative AI in certain domains. The cognitive limitations of human memory represent a significant disadvantage compared to Generative AI's extensive information storage and retrieval capabilities. Similarly, cognitive bias affects human judgment in ways that may be less prevalent in well-designed Generative AI systems.

### 3.5 Potential Future Directions

*3.5.1 Models of Human-Generative AI Interaction*

Based on this comparative analysis, three potential models of human-Generative AI interaction emerge for future decision-making in complex health services:

1. **Competition**: In domains where either humans or Generative AI demonstrate clear advantages without significant complementarity, competitive models may emerge. For example, Generative AI may increasingly outperform humans in policy tasks that require extensive repetition and/or pattern recognition across large datasets. Humans may lack the stamina and cognitive capacity to perform such tasks consistently.
2. **Cooperation**: In areas requiring diverse cognitive approaches, cooperation between humans and Generative AI may yield superior outcomes. For instance, policy planning might benefit from the contextual understanding and ethical judgment of humans, and the ability of Generative AI to process significant amounts of data.
3. **Convergence**: Over time, humans and Generative AI may converge in ways that cannot yet be foreseen or even converge to form a new entity.

*3.5.2 Implications for Workforce Development and System Design*

For humans to maintain meaningful roles in future decision-making, workforce development could focus on strengthening uniquely human attributes such as ethical reasoning, interpersonal skills, and contextual understanding. Educational programs for decision makers in complex health services could emphasize these distinctively human attributes, while also developing technical literacy to enable effective collaboration with Generative AI.

Simultaneously, health system designers could create decision architectures that leverage the complementary strengths of humans and Generative AI. This might involve developing Generative AI systems that explicitly incorporate human ethical frameworks, provide explanations for their recommendations to enable meaningful human oversight, and integrate smoothly into policy and regulatory workflows.

### 3.6 Potential Future Directions

This comparative analysis has several limitations. First, the attributes identified were based on literature reviews rather than direct empirical observation, potentially missing emerging attributes in rapidly evolving Generative AI systems. Second, the classification of attributes as helpful, detrimental, or context-dependent necessarily involves subjective judgment that may vary across different contexts in complex health services. Third, the analysis focused on attributes at a conceptual level without quantifying their relative importance or impact on decision quality.

There is an opportunity for future research to address these limitations:

- empirical studies of human-Generative AI collaboration in real-world settings;
- quantitative assessments of decision quality under different models of competition, cooperation, and convergence; and
- longitudinal studies tracking the evolution of Generative AI attributes and human-Generative AI relationships over time.

Additionally, research could explore how organizational and policy frameworks can best support ethical and effective collaboration between humans and Generative AI in complex health services.

## 4 CONCLUSIONS

This narrative comparison reveals that both humans and Generative AI possess predominantly helpful attributes, but with different patterns of strengths and weaknesses. Humans maintain their competitive advantage through unique attributes centered on ethical reasoning, interpersonal capabilities, contextual understanding, and metacognitive processes. Generative AI demonstrates strengths in information processing, pattern recognition, and memory, while facing challenges with hallucinations, lack of empathy, and lack of consideration for ethics, privacy, and human safety.

The complementary nature of these attribute profiles suggests that cooperation and convergence between humans and Generative AI will likely characterize future decision-making in complex health services, rather than pure competition or replacement. For humans to maintain meaningful roles in this future, workforce development could emphasize attributes unique to humans, while health system design could create architectures that effectively integrate the contributions of both humans and Generative AI.

As Reddy [11] noted, complex health services are already beginning to incorporate sophisticated Generative-AI-based decision support systems. This comparison suggests that an effective path forward may involve leveraging the unique strengths of both humans and Generative AI in complementary ways, rather than viewing the future as a zero-sum competition between different forms of higher intelligence. By understanding, cultivating, and incorporating the unique attributes that humans bring to decision-making, they may continue in meaningful decision-making roles in landscapes and environments increasingly influenced by Generative AI. There is a possibility that the attributes of these two entities may converge in future, or that the entities themselves may merge. However, it is unclear, at this point in time, whether convergence will occur and, if so, what shape and form it might take.

## 5 DATA AVAILABILITY

Data generated or analysed for this review are included in this article.

## 6 AUTHOR CONTRIBUTIONS

The first author extracted the data and wrote the manuscript. Both authors performed the analysis. The second author provided expert guidance throughout. Both authors read and approved the final manuscript.

## 7 CONFLICT OF INTEREST

The authors declare that they have no conflict of interest.